\documentclass[twocolumn,twoside,slac]{revtex4}
\usepackage{graphicx}
\usepackage{fancyhdr}
\begin{document}
\title{Asymmetric Errors}
\author{Roger Barlow}
\affiliation{Manchester University, UK and  Stanford University, USA}
\begin{abstract}
Errors quoted on results are often given in asymmetric form.
An account is given of the two ways these can arise
in an analysis, and the combination of 
asymmetric errors is discussed.  
It is shown that the usual method has no basis and is indeed wrong.
For asymmetric 
systematic errors, a consistent method is given, with detailed
examples.  For asymmetric statistical errors a
general approach is outlined.
\end{abstract}
\maketitle
\thispagestyle{fancy}
\font \csc=cmcsc12
\def \babar {{\csc BaBar}}

\section{Asymmetric Errors}
In the reporting of results from particle physics experiments it is common to
see values given with errors with different positive and negative numbers,
to denote a 68\% central confidence region which is not symmetric about the
central estimate. 
For example (one of many)  the Particle Data Group\cite{PDG} quote 
$$B.R. (f_2(1270) \to \pi\pi) = (84.7 ^{+2.4}_{-1.3}) \%.$$

The purpose of this note is 
to describe how such errors  arise  and how they
can properly be handled, particularly when two contributions are combined.
Current practice 
is to combine such errors separately, i.e. 
to add the $\sigma^+$ values together in quadrature, and then
do the same thing for the $\sigma^-$ values. This is not, to my knowledge,
documented anywhere and, as will be shown, is certainly wrong. 

There are two separate sources of asymmetry, which unfortunately 
require different treatments. 
We call these
`statistical' and `systematic'; the label is fairly
accurate though not entirely so, and they could
equally well be called `frequentist' and `Bayesian'. 

Asymmetric statistical errors arise when the 
log 
likelihood curve is not well described by a parabola~\cite{Eadie}.
The one sigma values (or, equivalently, the 68\% central confidence level
interval limits) are
read off the points at which $\ln L$ falls from its peak by ${1 \over 2}$ -- 
or, equivalently, when $\chi^2$ rises by 1.
This is not strictly accurate, and corrections 
should be made using Bartlett functions\cite{Frodesen}, but that lies 
beyond the scope of this note.

Asymmetric systematic errors arise when the
dependence of a result on a `nuisance parameter' is non-linear.
Because the dependence on such parameters -- theoretical values,
experimental calibration constants, and so forth --  is generally 
complicated, involving
Monte Carlo simulation, this study generally has to be performed by
evaluating the result $x$ at the
$-\sigma$ and $+\sigma$ values of the nuisance parameter $a$
(see~\cite{Durham} for a fuller account)
giving  $\sigma_x^-$
and $\sigma_x^+$.
($a \pm \sigma$ gives $\sigma_x^\pm$ or $\sigma_x^\mp$ according to the sign of
${dx \over da}$.)

This note summarises a 
full account of the procedure for asymmetric systematic errors which
can be found in~\cite{asymmetricpreprint}
and describes what has subsequently
been achieved for asymmetric statistical errors.
For
another critical account see~\cite{dAgostini}.

\section {Asymmetric Systematic Errors}

If $\sigma_x^-$ and $\sigma_x^+$ are 
different then this is 
a sign that the dependence of $x$ on $a$ is non-linear and the 
symmetric distribution in $a$
gives an asymmetric distribution in $x$.
In practice, if the difference is not large, one might be well 
advised to assume a straight line dependence and take the error as
symmetric, however we will assume that this is not a case where this is 
appropriate. 
We consider
cases where a non-linear effect is not small
enough to be ignored entirely, but not large enough to justify a 
long and intensive
investigation.
Such cases are common enough in practice.

\subsection {Models}

For simplicity we transform $a$ to the 
variable $u$ described by a unit Gaussian, and work with $X(u)=x(u)-x(0)$.
It is useful to define the mean $\sigma$, the difference $\alpha$, and the
asymmetry $A$:
\begin{equation}
\sigma = {\sigma^+ + \sigma^- \over 2}\qquad
\alpha = {\sigma^+ - \sigma^- \over 2}\qquad
A={\sigma^+ - \sigma^- \over \sigma^+ + \sigma^-}\label{Eq1}
\end{equation}
There are infinitely many non-linear
relationships between $u$ and $X$ that will go through the three determined
points.
We consider two. 
We make no claim that either of these is
`correct'. 
But working with asymmetric errors must involve 
some model of the non-linearity. 
Practitioners must select one of these two models, or some other 
(to which the same
formalism can be applied),  on the
basis of their knowledge of the problem, their preference and experience. 
 
\begin{itemize}
\item Model 1: Two straight lines

Two straight lines are drawn, meeting at the central value
\begin{eqnarray}
 X&=\sigma^+ u  \qquad u\geq 0 \nonumber \\
&=\sigma^- u  \qquad u \leq 0.
\end{eqnarray}

\item Model 2: A quadratic function 

The parabola through the three points is

\begin{equation}
X=\sigma u + \alpha u^2=\sigma u + A\sigma u^2.
\end{equation}

\end{itemize}

\begin{figure}[ht]
\centering
\includegraphics[width=50mm]{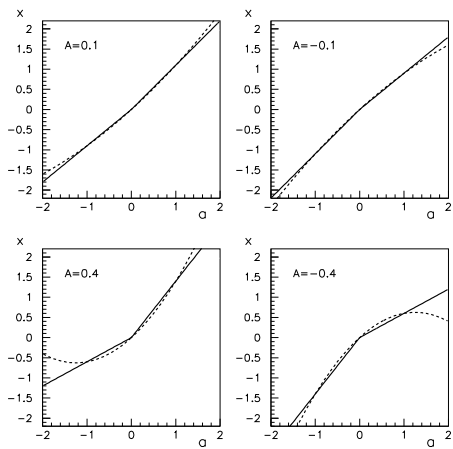}
\caption{ Some nonlinear dependencies \label{figmodels}}
\end{figure}

These forms are shown in Figure~\ref{figmodels} for a small asymmetry of 0.1, and 
a larger asymmetry of 0.4.

Model 1 is shown as a solid line, and Model 2 
is dashed.
Both go through the 3 specified points. The differences between them
within the range $-1\leq u \leq 1$ are not large; outside that range they
diverge considerably.

\begin{figure}[ht]
\centering
\includegraphics[width=50mm]{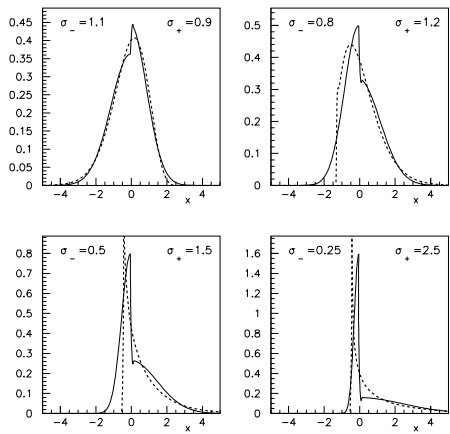}
\caption{ 
Probability Density Functions from Figure \ref{figmodels}
}
\label{figcom1}
\end{figure}

The distribution in $u$ is a unit Gaussian, $G(u)$, and the distribution 
in $X$ is obtained from
$P(X)={G(u) \over | dX/du |}$. 
Examples
are shown in Figure~\ref{figcom1}.
 For Model 1 (again a solid line) this gives a dimidated 
Gaussian -
two Gaussians with different standard deviation for $X>0$ and $X<0$.
This is sometimes called a `bifurcated Gaussian', but this 
is inaccurate. 
`Bifurcated' means `split' in the sense of forked. 
`Dimidated' means `cut in half', with the subsidiary meaning
of `having one part much smaller than the other'~\cite{OED}.
For Model 2 (dashed) with small asymmetries the curve is a distorted Gaussian,
given by ${G(u) \over |\sigma + 2 \alpha u |}$ with 
$u={\sqrt{\sigma^2 + 4 \alpha X} - \sigma \over 2 \alpha}$. For larger asymmetries and/or
larger $|X|$ values, the second root also has to be considered. 

It can be seen that the Model 1 dimidated Gaussian and Model 2 distorted Gaussian
are not dissimilar if the asymmetry is small, but are very different if the
asymmetry is large. 

\subsection {Bias}

If a nuisance parameter $u$ is distributed with a 
Gaussian probability distribution, and the quantity $X(u)$ is a 
nonlinear function of $u$, then the expectation $\langle X \rangle$
is not $X(\langle u \rangle )$.  

For model 1 one has \begin{equation}<X> = 
{\sigma^+ - \sigma^- \over \sqrt{2 \pi}}
\label{eqnbias1}
\end{equation}

For model 2 one has 
\begin{equation}
<X> = 
{\sigma^+ - \sigma^- \over 2 }  
 =\alpha 
\label{eqnbias2}
\end{equation}

Hence in these models, (or any others), 
if the result quoted is $X(0)$,
it is not the mean. It differs from it by an amount
of the order of the difference in the positive and negative errors.  
It is perhaps defensible
as a number to quote as the result as 
it is still the median - there is a 50\% chance that the true value is
below it and a 50\% chance that it is above.

\subsection {Adding Errors}

If a derived quantity $z$ contains parts from two quantities $x$ and 
$y$, so that $z=x+y$, the distribution in $z$ is given by the convolution:

\begin{equation}
f_z(z)=\int dx f_x(x) f_y(z-x)
\label{eqnConvolute}
\end{equation}

\begin{figure}[ht]
\centering
\includegraphics[width=50mm]{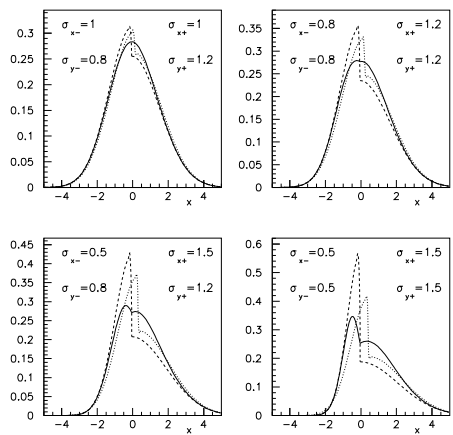}
\caption{ 
 Examples of the distributions from combined asymmetric
errors using Model 1. 
}
\label{figcom2}
\end{figure}

With Model 1
the convolution can be done analytically. Some 
results
for typical cases 
 are shown in Figure~\ref{figcom2}.
The
solid line shows the convolution, the dashed line 
is obtained by adding the positive and negative standard deviations separately
in quadrature (the `usual procedure').
The dotted line is described later.

The solid and dashed curves disagree markedly.
The `usual procedure' curve  has a larger skew than the 
convolution. 
This is obvious.
If two distributions with the same asymmetry are added
the `usual procedure' will give a distribution just scaled by $\sqrt 2$,
 with the same asymmetry.
This violates the Central Limit Theorem, which says that convoluting
identical distributions
must result in a combined distribution which is more Gaussian, and therefore
more symmetric,  than its components. This shows that the
`usual procedure' for adding asymmetric errors is inconsistent.  

\subsection {A consistent addition technique}

If a distribution for $x$ is described by some function,
$f(x;x_0,\sigma^+,\sigma^-)$, which is a Gaussian transformed according to
 Model 1 or Model 2 or anything else, then `combination of errors' involves
a convolution of two such functions according to Equation~\ref{eqnConvolute}.
This combined function is not necessarily a function of the same form: it is
a special property of the Gaussian that the convolution of two 
Gaussians gives a third.
The (solid line)  convolution of two
dimidated Gaussians is not itself
a dimidated Gaussian.
Figure~\ref{figcom2} is a demonstration of this. \

Although the form of the function is changed by a convolution, some things
are preserved. 
The semi-invariant cumulants of Thi\` ele (the coefficients of the power series expansion of the 
log of the Fourier Transform) add under convolution. The first two of these
are
the usual mean and variance. The third is the unnormalised skew:
\begin{equation}
\gamma = <x^3> - 3<x><x^2> + 2 <x>^3
\label{eqnskew}
\end{equation}
Within the context of any model, a consistent approach to the combination
of
errors
is to find the mean, variance and skew: $\mu$, $V$ and $\gamma$, 
for each 
contributing function separately.  Adding these up gives the
mean, variance and skew of the combined function. Working within the model
one then determines the values of
$\sigma_-, \sigma_+$, and $x_0$ 
that give this mean, variance and skew.

\subsection {Model 1}

For Model 1, for which
$\langle x^3 \rangle ={2 \over \sqrt{2 \pi}} (\sigma_+^3 - \sigma_-^3)$ we
have
\begin{eqnarray}
&\mu=x_0+{1 \over \sqrt{2 \pi}} 
(\sigma^+ - \sigma^-)\nonumber\\
&V= 
 \sigma^2 + \alpha^2\left( 1-{2 \over  \pi}\right)\nonumber\\
&\gamma=
{1 \over \sqrt{2 \pi}} \big[
2
({\sigma^+}^3 - {\sigma^-}^3) 
-{3 \over 2}
({\sigma^+} - {\sigma^-}) 
({\sigma^+}^2 + {\sigma^-}^2) 
\nonumber\\
&+{1 \over \pi}
({\sigma^+} - {\sigma^-}) ^3
\big]
\label{eqnDict1}
\end{eqnarray}
Given several error contributions 
the Equations~\ref{eqnDict1}  give the cumulants $\mu$, $V$ and $\gamma$ 
of each.  Adding these up gives 
the first three cumulants  of the combined distribution.
Then one can find the set of
parameters
 $\sigma^-, \sigma^+, x_0$ 
which give these values by using Equations~\ref{eqnDict1}  in the other sense.

It is convenient to work with 
$\Delta$,
where $\Delta$ is the difference between the final $x_0$ and the
sum of the individual ones.
The parameter 
is needed because of the bias mentioned earlier.
Even though each contribution may have $x_0=0$, i.e. it describes a spread about the
quoted result, it has non-zero $\mu_i$ through the bias  effect
(c.f. Equations ~\ref{eqnbias1} and \ref{eqnbias2} ).
The $\sigma^+$ and $\sigma^-$ of the combined distribution, obtained from 
the total $V$ and $\gamma$, will in general not give the right $\mu$ unless 
a location shift $\Delta$ is added. {\it The value of the quoted result will shift.}

Recalling section B, for the original distribution
one could defend quoting the  central value as it 
was the median, even though it was not the mean. 
The convoluted distribution not only has a non-zero mean, it also
(as can be seen in Figure~\ref{figcom2} ) has non-zero median.
If you want to combine asymmetric errors 
then you have to accept that the quoted value will
shift.  To make this correction requires a real belief in the 
asymmetry of the error 
values. 
At this point practitioners,
unless they are sure that their errors really do have a
significant asymmetry,
may be
persuaded to revert to quoting symmetric errors.

Solving the Equations~\ref{eqnDict1}   for 
 $\sigma^-, \sigma^+$ and $ x_0$ 
given 
$\mu$, $V$ and $\gamma$ 
has to be done numerically.  
A program for this is available on \hfill {\tt http://www.slac.stanford.edu/$\sim$barlow}.
Some results are shown in the dotted curve of Figure~\ref{figcom2} and Table 1.

\begin{table}[ht]
\begin{center}
\caption{Adding errors in Model 1}
\begin{tabular}{|cccc|ccc|}
\hline 
$\sigma_x^-$ &
$\sigma_x^+ $&
$\sigma_y^- $&
$\sigma_y^+ $&
$\sigma^{-} $&
$\sigma^{+} $&
$\Delta$\\
\hline 
1.0 & 1.0 & 0.8 & 1.2 & 1.32 & 1.52 & 0.08 \\
0.8 & 1.2 & 0.8 & 1.2 & 1.22 & 1.61 & 0.16 \\
0.5 & 1.5 & 0.8 & 1.2 & 1.09 & 1.78 & 0.28 \\
0.5 & 1.5 & 0.5 & 1.5 & 0.97 & 1.93 & 0.41 \\
\hline
\end{tabular}
\label{table1}
\end{center}
\end{table}

It is apparent that the dotted curve
agrees much better with the solid one than the
`usual procedure' dashed curve does.
It is not an exact match, but 
does an acceptable job given that there are only 3 adjustable parameters in the function.
If the shape of the solid curve is to be represented by a dimidated Gaussian,
then it is plausible that the dotted curve is the `best' such representation.

\subsection {Model 2}

The equivalent of Equations~\ref{eqnDict1} are
\begin{eqnarray}
&\mu=x_0+\alpha\nonumber\\
&V=\sigma^2 + 2\alpha^2\nonumber\\
&\gamma=6\sigma^2\alpha + 8 \alpha^3
\label{eqnDict2}
\end{eqnarray}

As with Method 1, these are used to find the cumulants of each contributing distribution,
which are summed to give the three totals, and then Equation~\ref{eqnDict2} is 
used again to find the
parameters of the distorted Gaussian with this mean, variance and skew.
The web program will also do these calculations

\begin{figure}[ht]
\centering
\includegraphics[width=50mm]{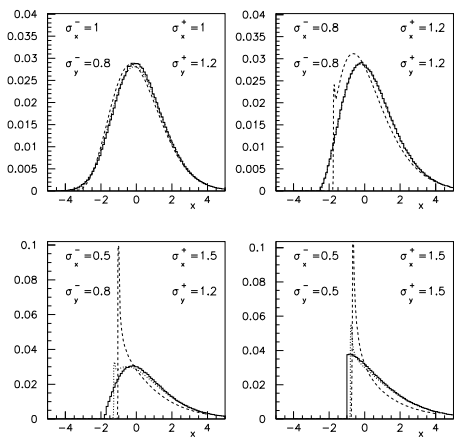}
\caption{ 
 Examples of combined
errors using Model 2. 
\label{figcom3}}
\end{figure}

Some results are shown in Figure~\ref{figcom3} and Table~\ref{table2}. The true convolution
cannot be done analytically but can be done by a Monte Carlo calculation.
\begin{table}[ht]
\begin{center}
\caption{Adding errors in Model 2}
\begin{tabular}{|cccc|ccc|}
\hline 
$\sigma_x^-$ &
$\sigma_x^+ $&
$\sigma_y^- $&
$\sigma_y^+ $&
$\sigma^{-} $&
$\sigma^{+} $&
$\Delta$\\
\hline 
1.0 & 1.0 & 0.8 & 1.2 & 1.33 & 1.54 & 0.10 \\
0.8 & 1.2 & 0.8 & 1.2 & 1.25 & 1.64 & 0.20 \\
0.5 & 1.5 & 0.8 & 1.2 & 1.12 & 1.88 & 0.35 \\
0.5 & 1.5 & 0.5 & 1.5 & 1.13 & 2.07 & 0.53 \\
\hline
\end{tabular}
\label{table2}
\end{center}
\end{table}

Again the true curves (solid) are not well reproduced by the `usual procedure' (dashed)
but the curves with the correct cumulants (dotted) do a 
good job. 
(The sharp behaviour at the edge of the curves is 
due to the turning point of the parabola.)

\subsection {Evaluating $\chi^2$}

For Model 1 the $\chi^2$ contribution from a discrepancy $\delta$ is 
just $\delta^2/{\sigma^+}^2$ or $\delta^2/{\sigma^-}^2$ as appropriate.  
This is
manifestly inelegant, especially for minimisation procedures as the value goes through zero. 

For Model 2 one has
\begin{equation}
\delta=\sigma  u + A \sigma u^2.
\end{equation}

This can be considered as a quadratic for $u$ with solution
which when squared gives
$u^2$, the $\chi^2$ contribution, as
\begin{equation}
u^2={2+4A{\delta \over \sigma} -2 (1+4A{\delta \over \sigma})^{1 \over 2}\over 4 A^2}
\end{equation}
This is not really exact, in that it only takes one 
branch of the solution, the one approximating to
the straight line, 
and does not consider the extra possibility that the $\delta$ value could come
from an improbable $u$ value the other side of the turning point of the parabola.
Given this imperfection it makes sense to 
expand the square root as a Taylor series, which, neglecting correction
terms above 
the second power, leads to
\begin{equation}
\chi^2=({\delta \over \sigma})^2 \left(1-2A ({\delta \over \sigma})+
5 A^2 ({\delta \over \sigma})^2\right).\end{equation}

This provides a sensible form for $\chi^2$ from asymmetric errors.
It is important to keep the $\delta^4$ term rather than stopping at $\delta^3$
to ensure $\chi^2$ stays positive!  Adding higher orders does not
have a great effect.
We recommend it for consideration when it is required
(e.g. in fitting parton distribution functions)
to form
a  $\chi^2$ from asymmetric errors

\subsection {Weighted means}

The `best' estimate (i.e. unbiassed and with
smallest variance) from several measurements $x_i$ 
with different (symmetric) errors $\sigma_i$
is given by a weighted sum
with $w_i = 1/\sigma_i^2$. We wish to find the equivalent for asymmetric
errors. 

As noted earlier, when sampling from an asymmetric distribution the 
result is biassed
towards the tail. The expectation value $\langle x \rangle$
is not the location parameter $x$. 
So for an unbiassed estimator one must take
\begin{equation}
\hat x= \sum w_i(x_i-b_i) / \sum w_i
\end{equation}
where
\begin{equation}
b={\sigma^+-\sigma^- \over \sqrt{2 \pi}} \quad \hbox{(Model 1)} \qquad 
b=\alpha \quad \hbox{(Model 2)}
\end{equation}
The variance of this is given by
\begin{equation}
V={\sum w_i^2 V_i \over \left( \sum w_i \right)^2}
\end{equation}
where $V_i$ is the variance of the $i^{th}$  measurement about its mean.
Differentiating with respect to $w_i$ to find the minimum gives
\begin{equation}
{2 w_i V_i \over \left( \sum w_j\right)^2} - {2 \sum w_j^2 V_j \over \left( \sum w_j \right)^3}=0\qquad \forall i 
\end{equation}
which is satisfied by $w_i = 1/V_i$.
This is the equivalent of the 
familiar weighting by $1/\sigma^2$. 
The weights are given, depending on the Model,
 by (see Equations~\ref{eqnDict1} and \ref{eqnDict2})
\begin{equation}
V=\sigma^2 +(1-{2\over  \pi})\alpha^2 \qquad \hbox{or}\qquad V=\sigma^2 + 2 \alpha^2
\end{equation}

Note that this is not the Maximum Liklelihood estimator - 
writing down the likelihood in terms
of the $\chi^2$ and differentiating does not give a nice form -
so in principle there may be better estimators, but they will not have the simple form of a weighted sum.

\section {Asymmetric Statistical Errors}

 As explained earlier, (log) likelihood curves are
used to obtain the maximum likelihood estimate for a parameter and
also the 68\% central interval -- taken as the values at which
$\ln L$ falls by ${1 \over 2}$ from its peak.  
For large $N$ this curve is a parabola, but for finite $N$
it is generally asymmetric, and the two points are not
equidistant about the peak.

The bias, if any, 
is not connected to the form of the curve, which is a likelihood and not a pdf.
Evaluating a bias is done by integrating over the measured value not the
theoretical parameter.  We will assume for simplicity that these
estimates are bias free. This means that when combining errors there will be no
shift of the quoted value.

\subsection{Combining asymmetric statistical errors}

Suppose estimates $\hat a$ and $\hat b$ are obtained by this method for
variables $a$ and $b$. 
$a$ could typically be an
estimate of the total number of events in a signal region, and $b$
the (scaled and negated) estimate of background, obtained from a sideband. 
We are interested in $u=a+b$,
taking $\hat u=\hat a+\hat b$. What are the errors to be quoted on $\hat u$?

\subsection {Likelihood functions known}

\label{fullsection}
We first consider the case where the likelihood functions 
$L_a(\vec x|a)$ and $L_b(\vec x|b)$ are given. 

For the symmetric Gaussian  case, 
the answer is
well known. Suppose that the likelihoods are both Gaussian, and further
that $\sigma_a=\sigma_b=\sigma$. 
The log likelihood term
\begin{equation}
\left( {\hat a - a \over \sigma} \right)^2 
+ \left( {\hat b - b \over \sigma} \right)^2 
\end{equation}
can be rewritten
\begin{equation}
{1 \over 2} 
\left(
{\hat a+\hat b - (a+b) \over \sigma} \right)^2 
+{1 \over 2} 
\left(
{\hat a-\hat b - (a-b) \over \sigma} \right)^2 
\end{equation}
so the likelihood is the product of Gaussians for $u=a+b$ and
$v=a-b$, with standard deviations $\sqrt 2 \sigma$.

Picking a particular value of $v$, one can then trivially construct 
the 68\% confidence region for $u$ as 
$[\hat u - \sqrt 2 \sigma,\hat u + \sqrt 2 \sigma]$.
Picking another value of $v$, indeed any other value  of $v$,
one obtains the same region for $u$.  We can therefore say with
68\% confidence that these limits enclose the true value of $u$, 
whatever the value of $v$. 
The uninteresting part of $a$ and $b$ has been `parametrised away'.
This is, of course, the standard result from the combination of errors formula,
but derived in a 
frequentist way using Neyman-style confidence intervals.
We could construct the limits on $u$ by finding $\hat u+\sigma_u^+$
such that
the integrated probability of a result as small as or smaller 
than the data be 16\%, and similarly for $\sigma_u^-$,  rather than taking the 
$\Delta \ln L=-{1 \over 2}$ shortcut, and it would not affect the
argument. 

The question now is how to generalise this. 
For this to be possible the likelihood must factorise
\begin{equation}
L(\vec x|a,b)=L_u(\vec x|u) L_v(\vec x|v)
\end{equation}
with a suitable choice of the parameter $v$ and the functions $L_u$ and $L_v$.
Then we can use the same argument:  for any value of $v$ 
the limits on $u$ are the same, depending  only on $L_u(\vec x|u)$. 
Because they are true for any $v$ they are true for all $v$, and 
thus in general.

There are cases where this can clearly be done.
For two Gaussians  with  $\sigma_a \neq \sigma_b$
the result is the same as above but with 
$v=a \sigma_b^2 -b \sigma_a^2 $. For two Poisson
distributions $v$ is $a/b$. There are cases 
(with multiple peaks)
where 
it cannot be done, but let us hope that these are artificially pathological. 

On the basis that if it cannot be done, the question is
unanswerable, let us  assume that it is possible in the case
being studied, and see how far we can proceed.
Finding the form of $v$ is liable to be difficult, and as it is not
actually used in the answer we would like to avoid doing so.
The limits on $u$ are read off from the $\Delta \ln L(\vec x|u,v) = -{1 \over 2}$
points where $v$ can have any value provided it is fixed.
Let us choose
$v=\hat v$, the value at the peak.  This is the value of $v$
at which $L_v(v)$ is a maximum.  Hence when we
consider any other value of $u$, we can find $v=\hat v$
by finding the point at which the likelihood
is a maximum,
varying $a-b$, or $a$, or 
$b$, or any other combination, always keeping $a+b$ fixed.
We can read the limits off a 1 dimensional plot of $\ln L_{max}(\vec x|u)$,
where the `max' suffix denotes that at each value of $u$ we search the subspace to pick out the maximum value.

This generalises to more complicated situations. If $u=a+b+c$ we again scan
the $\ln L_{max}(\vec x|u)$ function, where the subspace is now 2 dimensional.

\subsection {Likelihood functions not completely known}

\begin{figure}[ht]
\centering
\includegraphics[width=40mm]{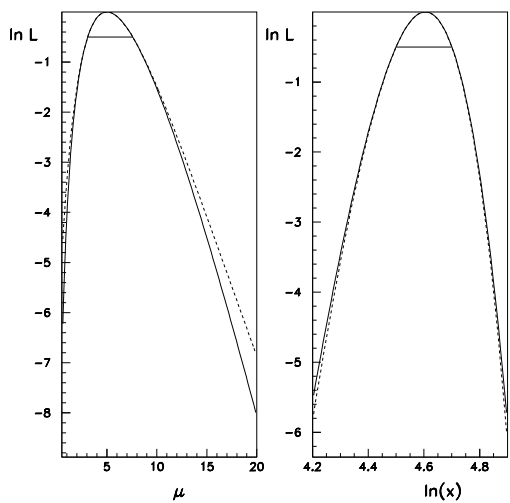}
\caption{ Approximations using Equation~\ref{parametrise}} \label{figparab}
\end{figure}

In many cases the likelihood functions for $a$ and $b$ will not be
given, merely
estimates $\hat a$ and $\hat b$ and 
their asymmetric errors 
$\sigma^+_a$,
$\sigma^-_a$,
$\sigma^+_b$ and
$\sigma^-_b$. All we can do is to use these to provide
best guess functions $L_a(\vec x|a)$ and $L_b(\vec x|b)$.
A parametrisation of suitable shapes, which 
for $\sigma^+ \sim \sigma^-$ approximate to a parabola,
must be provided. 
Choosing a suitable parametrisation is
not trivial. The obvious choice of introducing small higher-order terms fails
as these dominate far from the peak.
A likely candidate is:
\begin{equation}
\ln{ L}(a)=-{1 \over 2} \left( \ln{\left(1+a/\gamma \right)}\over \ln{\beta} \right)^2
\label{parametrise}
\end{equation}
where $\beta = {\sigma_+ /\sigma_-}$ and $\gamma={\sigma_+ \sigma_- \over \sigma_+ - \sigma_-}$.
This describes the usual parabola, but with the x-axis stretched by
an amount that changes linearly with distance.
Figure~\ref{figparab} shows two illustrative results.
 The first 
is the Poisson likelihood from 5 observed events (solid line) for which the 
estimate using the $\Delta \ln L={1 \over 2}$ points is
$\mu = 5^{+2.58}_{-1.92}$, as shown.  The dashed line is that obtained 
inserting these numbers into Equation~\ref{parametrise}.
The second considers a measurement of $x=100 \pm 10$, of which the logarithm has been
taken, to give a value $4.605^{+0.095}_{-0.105}$. 
Again, the solid line is the true curve and the dashed line the parametrisation.
In both cases 
the agreement is
excellent over the range $\approx \pm 1 \sigma$ and reasonable over the
range 
$\approx \pm 3 \sigma$.

To check the correctness of the method we can use
the combination of two Poisson numbers, for which the result is 
known.
First indications are that the errors obtained from the parametrisation are indeed 
closer to the true Poisson errors than those obtained from the
usual  technique.

\subsection {Combination of Results}

A related problem is to find the combined estimate
$\hat u$ given estimates $\hat a$ and $\hat b$ (which have asymmetric errors).  Here $a$ and $b$ could be results from different 
channels or different experiments.
This can be regarded as a special case, constrained to $a=b$, i.e. $v=0$, but 
this is rather contrived. It is more direct just to say that one uses the log likelihood 
which is the sum of the two separate functions, and determines the peak and the 
$\Delta \ln L=-{1 \over 2}$  points from that.  If the functions are known this is unproblematic, if
only the errors are given then the same parametrisation technique 
can be used.

\section {Conclusions}

If asymmetric errrors  cannot be avoided they need careful
handling.

A method is suggested and a program
provided for combining asymmetric systematic errors. It
is not `rigorously correct' but such perfection is impossible. 
Unlike the usual method, 
it is at least 
open about its assumptions and mathematically consistent.

Formul\ae\ for $\chi^2$ and weighted sums are given.

A method is proposed for combining asymmetric statistical errors
if the likelihood functions 
are known.
Work is in progress to enable it to be used given
 only the results and their errors.

\begin{acknowledgments}

The author a gratefully acknowledges the support of the Fulbright Foundation.

\end{acknowledgments}


\begin{thebibliography}{9}   
\bibitem{PDG} D.E. Groom {\it et al.\/}, Eur. Phys. J. {\bf C15} 1 (2000).

\bibitem{Eadie}  W. T. Eadie et al, ``Statistical Methods in Experimental Physics'', 
North Holland, 1971.

\bibitem{Frodesen} A.G. Frodesen {\it et al.\/} ``Probablity and Statistics in Particle Physics'', Universitetsforlaget Bergen-Oslo-Tromso (1979), pp 236-239.



\bibitem{Durham} R. J. Barlow ``Systematic Errors: Facts and Fictions" in Proc. Durham 
conference on Advanced Statistical Techniques in Particle Physics, 
M. R. Whalley and L. Lyons (Eds). IPPP/02/39. 2002.

\bibitem{asymmetricpreprint}
 R. J. Barlow, ``Asymmetric Systematic Errors'' preprint MAN/HEP/03/02, 
ArXiv:physics/030613.

\bibitem{dAgostini} G. D'Agostini ``Bayesian Reasoning in Data Analysis: a Critical Guide'', World Scientific (2003).

\bibitem{OED} 
The Shorter Oxford 
English Dictionary, Vol I (A-M) p 190 and p 551 of the 3rd edition (1977). 

\end{thebibliography}
\end{document}